\documentclass[11pt]{article}
 \usepackage{graphicx}
\begin{document}

{\scriptsize \em
13th. April 2016}

\begin{center}
{\large{\bf Comment on two papers claiming records for the lowest carrier
concentration at which superconductivity has been observed}}
\end{center}

\vskip 2em

\begin{center}
{D.M. Eagles}$^{*}$
\end{center}

\vskip 1em
\begin{center}
{\em 19 Holt Road, Harold Hill, Romford, Essex RM3 8PN, England} 
\end{center}

\begin{center}
\end{center}

\begin{center}
\end{center}

A 2013 paper \cite{Li13} and a recent preprint \cite{Pr16} claim records
for the lowest carrier concentrations at which superconductivity has
been observed, at concentrations of 5.5$\times 10^{17}$ cm$^{-3}$ in a
crystal of SrTiO$_{3}$ \cite{Li13} and 3$\times 10^{17}$ cm$^{-3}$ in a
crystal of Bi \cite{Pr16}.   However, evidence for superconductivity at
much lower carrier concentrations at low-temperatures, of the order of
10$^{15}$ cm$^{-3}$, was obtained by myself and colleagues 
in the late 1980's \cite{Ta86,Ea86,Ea89} in a reduced ceramic sample of
SrTiO$_3$ with 3\% of Ti replaced by Zr. The Zr does not produce any
charge carriers, but does produce large increases in their effective
mass \cite{Ea69a}, as shown by analysis of magnetic-field penetration
depth results of Hulm et al \cite{Hu67}.  The increase in mass appears
to be sufficient to increase the electron-phonon coupling above the
threshold at which pairing without superconductivity at very low carrier
concentrations is possible \cite{Ea69b}.

We were unable to measure the low-temperature Hall coefficients in
the ceramic samples studied, and so had to infer the low-temperature
carrier concentrations, $n$, indirectly.  However, two different methods
\cite{Ea86,Ea89} gave results in agreement with each other.  The first
method was based on inferences from room-temperature Hall measurements
and resistivities as a function of temperature about which side of the
maximum in $T_c$ versus carrier concentration the six superconducting
samples lay, and approximately where, and led to an estimate that
3$\times 10^{14}$ cm$^{-3} < n < 5\times 10^{15}$ cm$^{-3}$  for the
sample mentioned.  The second method was based on detailed analysis of
the resistivity between 4.2 K and 75 mK, and resulted in an estimate of
$n\approx 3\times 10^{14}$ cm $^{-3}$, at the lower end of the previous
estimate. There was a temperature range between 130 mK and 75 mK in which
pairing without superconductivity raised the resistance above what it
would have been with no pairing.  These low carrier concentrations and
analysis showed that we were in the Bose-Einstein condensation r\'egime
in this sample.  The results were in qualitative, but not quantitative,
agreement to predictions made by me in 1969 \cite{Ea69b}.

Later we examined two macrocrystalline samples \cite{An90}, but in these
samples some non-uniform states appeared to prevent our reaching the
Bose-condensation regime.  We did find a sharp drop in resistance at
about 600 mK in both samples, one of which had an average concentration
at 2.3 K of about  $5\times 10^{15}$ cm$^{-3}$.  The drops in resistance
were probably due to small superconducting parts of the samples with
much higher $n$ than the average over the whole.  I conjecture that
the difference between the ceramic and macrocrystalline samples may be
connected with a larger effect of defects and disorder on charge density
waves than on superconductivity.

While I appreciate the careful studies of single crystals in
\cite{Li13,Pr16}, I  was disappointed that, at the time of writing of
their papers, the authors were apparently unaware of our earlier work
indicating superconductivity at much lower carrier concentrations in a
ceramic sample of Zr-doped SrTO$_3$.

\vskip 1em
\noindent
$^{*}$ E-mail: d.eagles@ic.ac.uk\newline

\end{document}